\documentclass[pra,preprint,tightenlines,showpacs]{revtex4}
\usepackage{amssymb}

\usepackage{bm,dcolumn,graphicx}

\begin{document}

\title{Hyperfine quenching of the metastable $^{3}\!P_{0,2}$
states in divalent atoms}

\author{Sergey G. Porsev}
\altaffiliation{Permanent Address: Petersburg Nuclear Physics
Institute, Gatchina, Leningrad district, 188300, Russia.}
\affiliation{Physics Department, University of Nevada, Reno,
Nevada 89557-0058.}
\author{Andrei Derevianko}
%\email{andrei@physics.unr.edu}
\affiliation{Physics Department, University of Nevada, Reno,
Nevada 89557-0058.}

\date{\today}

\begin{abstract}
Hyperfine quenching rates of the lowest-energy metastable
$^{3}\!P_{0}$ and $^{3}\!P_{2}$ states of Mg, Ca, Sr, and Yb atoms
are computed. The calculations are carried out using
{\em ab initio} relativistic many-body methods. The computed
lifetimes may be useful for designing  novel ultra-precise
optical clocks
and trapping experiments with the $^{3}\!P_{2}$ fermionic
isotopes. The resulting natural widths of the $^3\!P_0-^1\!S_0$
clock transition  are 0.44 mHz for $^{25}$Mg,
2.2 mHz for $^{43}$Ca, 7.6 mHz for $^{87}$Sr, 43.5 mHz for
$^{171}$Yb, and 38.5 mHz for $^{173}$Yb. Compared to
the bosonic isotopes, the lifetime of the
$^{3}\!P_{2}$ states in fermionic isotopes is noticeably shortened
by the hyperfine quenching but still remains long enough for trapping
experiments.
\end{abstract}

%32.70.Cs Oscillator strengths, lifetimes, transition moments
%32.10.Fn Fine and hyperfine structure
%31.15.Ar Ab initio calculations
\maketitle

%=====================
\section{Introduction}
%=====================

\begin{figure}[h]
\begin{center}
\includegraphics*[scale=1]{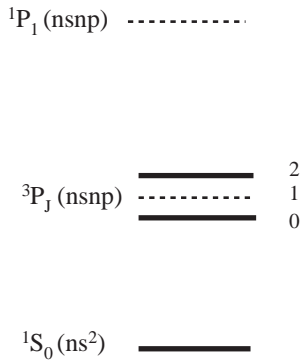}
\end{center}
\caption{ Lowest-lying energy levels of Mg ($n$=3), Ca ($n$=4), Sr
($n$=5), and Yb ($n$=6), relevant to the radiative decay of the
$nsnp\,^3\!P_{0,2}$ states. The hyperfine quenching predominantly is caused
by the admixture of the  $nsnp\,^3\!P_1$ and $nsnp\,^1\!P_1$ states.}
\label{Fig:Levels}
\end{figure}

This work is motivated by  emerging experiments  with cold
divalent atoms Mg, Ca, Sr and Yb~\cite{CopenhagenWorkshop2003}.
For example, the recently attained Bose-Einstein condensate of the
ground-state Yb \cite{TakMakKom03} may offer new insights into the
physics of degenerate quantum gases due to a vast number of
available isotopes and relative simplicity of molecular
potentials. As to the $^3\!P_2$  metastable states (see
Fig.~\ref{Fig:Levels}), it was realized that the non-scalar nature
of the $^3\!P_2$ states may be used to overcome the unfeasibility
of magnetic trapping of the  spherically-symmetric $^1\!S_0$
ground states~\cite{KatIdoIso00,NagSimLah03,XuLofHal03,GruHem02}.
Knowing  radiative lifetimes of the other, $^3\!P_0$, metastable
states is required in developing the next generation of
ultraprecise optical atomic
clocks~\cite{KatTakPal03,CouQueKov03,ParYoo03,PorDerFor03}. Here
the clockwork is  based on cold atoms confined to sites of an
engineered optical lattice. The lifetime determines the natural
width of the clock transition between the ground and the $^3\!P_0$
state.

For all {\em bosonic} isotopes of Mg, Ca, Sr, and Yb, the nuclear
spin $I$ vanishes and these isotopes lack hyperfine structure. For
bosonic isotopes the $^3\!P_0$ state may decay only via very weak
multi-photon (e.g., E1-M1) transitions. However, for {\em
fermionic} isotopes (Table~\ref{Tab:NucParams}), $I \neq 0$, a new
radiative decay channel becomes available due to the hyperfine
interaction (HFI). The HFI, although small, admixes atomic levels
of the total angular momentum $J=1$ thus opening an
electric-dipole branch to the ground state. The resulting
HFI-induced E1 decays do determine the lifetimes of the $^3\!P_0$
states. As to the  $^3\!P_2$ states, here the single-photon decays
are allowed, but being of non-E1 character, are very weak. The
lifetimes are long and range from  15 seconds for Yb to 2 hours
for Ca~\cite{PorRakKoz99P,Der01}. As we demonstrate here,
depending on an isotope, the hyperfine quenching of the  $^3\!P_2$
states  is either comparable to or is much faster than the small
non-E1 rates.

\begin{table}[h]
\caption{ Nuclear parameters of the stable fermionic isotopes of
Mg, Ca, Sr, and Yb. Here $I$ are the nuclear spins and
$\mu/\mu_N $ are the nuclear magnetic moments expressed in
units of the nuclear magneton $\mu_N$. }
\label{Tab:NucParams}
\begin{ruledtabular}
\begin{tabular}{ccc}
Isotope & $I$ & $\mu /\mu _{N}$  \\
\hline
$^{25}$Mg & 5/2 & -0.85546  \\
$^{43}$Ca & 7/2 & -1.31727  \\
$^{87}$Sr & 9/2 & -1.09283  \\
$^{171}$Yb & 1/2 & 0.4919   \\
$^{173}$Yb & 5/2 & -0.6776  \\
\end{tabular}
\end{ruledtabular}
\end{table}

A detailed theoretical analysis of the hyperfine quenching has
been limited so far to astrophysically-important He, Be and Mg and
their isoelectronic
sequences~\cite{Gar62,JohChePla97,BraJudAbo98,BraJudPro02}. The
hyperfine quenching of the $^3P_0$ states of Sr and Yb has been
estimated in Refs.~\cite{KatTakPal03,PorDerFor03}. Here we carry
out {\em ab initio} relativistic many-body atomic structure
calculations to extend and refine these previous studies. We find
that the resulting natural widths of the $^3\!P_0 - ^1\!S_0$
 clock transition  are 0.44 mHz for $^{25}$Mg,
2.2 mHz for $^{43}$Ca, 7.6 mHz for $^{87}$Sr, 43.5 mHz for
$^{171}$Yb, and 38.5 mHz for $^{173}$Yb. Compared to
the bosonic isotopes, the lifetime of the
$^{3}\!P_{2}$ states in fermionic isotopes is noticeably  shortened
by the hyperfine quenching but still remains long enough for trapping
experiments.

The paper is organized as follows. First, in
Section~\ref{Sec:Derivation} we derive the hyperfine quenching
rates using perturbation theory. The solution of many-body atomic
problem and numerical details are given in
Section~\ref{Sec:Method}. Finally, we present the results, compare
with the previous calculations, and draw the conclusions in
Section~\ref{Sec:Conclusions}. Unless noted otherwise, atomic
units ($\hbar=|e|=m_e\equiv 1$) are used throughout.

%================================================
\section{Derivation of hyperfine quenching rates}
%================================================
\label{Sec:Derivation}
In the presence of nuclear moments, the
total electronic angular momentum $J$ no longer remains a good
quantum number. The atomic energy levels are characterized instead
by the total angular momentum $\mathbf{F}=\mathbf{J}+\mathbf{I}$.
Nevertheless, the coupling of the electronic and the nuclear
momenta is small and in this section we employ the first-order
perturbation theory in the magnetic-dipole hyperfine interaction
to compute the modified atomic wave functions of the $^{3}P_{0;2}$
levels. With these perturbed wave functions the hyperfine
quenching rates are obtained with the conventional Fermi golden
rule.

Before proceeding with the outlined derivation, we notice that in
this problem there are two types of perturbations: the hyperfine
interaction and the interaction with the electromagnetic field.
Here we treat the HFI as the dominant interaction, determine the
hyperfine structure first, and as the next step compute the
lifetimes. This approach is valid as long as the radiative width
of the $^3\!P_1$ level is much smaller than the fine-structure
intervals between the components of the  $^3\!P_J$
multiplet~\cite{JohChePla97}. We verified that this inequality
holds for all the atoms under consideration.

We develop the formalism in terms of the hyperfine states $|\gamma
(IJ) F M_{F}\rangle$. Here the angular momenta $I$ and $J$ are
conventionally coupled to produce a state of definite total
momentum $F$ and its projection $M_F$, and $\gamma$ encapsulates
all other atomic quantum numbers. In the first order of
perturbation theory in the hyperfine interaction,
$H_{\mathrm{HFI}}$, the correction to the hyperfine sub-level
$|\gamma (IJ) FM_{F}\rangle$ of the metastable state $|\gamma
J\rangle$ reads
%===========================================================================
\begin{equation}
|\gamma (IJ) FM_{F}\rangle ^{\left( 1\right) } =
\sum_{\gamma^{\prime }J^{\prime
}}|\gamma^{\prime }(IJ^{\prime }) FM_{F}\rangle
\frac{\langle \gamma^{\prime}(IJ^{\prime })FM_{F}|H_{\mathrm{HFI}}|\gamma (IJ)FM_{F}\rangle }
{E\left( \gamma^{\prime }J^{\prime
}\right) -E\left( \gamma J\right) } \, ,
\label{Eq:WFCorr}
\end{equation}
%===========================================================================
where $E\left( \gamma J\right)$ are the energies of atomic states.
In the above expression, we have taken into account that
$H_{\mathrm{HFI}}$ is a scalar, so the total angular momentum $F$
and its projection $M_F$ are conserved. In general, the hyperfine
coupling Hamiltonian, $H_\mathrm{HFI}$,
may be represented as a sum over
multipole nuclear moments $\mathcal{M}^{(k)}$ of rank $k$ combined
with the even-parity electronic
coupling operators $\mathcal{T}^{(k)}$ of the same rank so
that the total interaction is rotationally and P-- invariant.
For the hyperfine quenching of the $J=0$ states via couplings to rapidly decaying
$J=1$ states, we may truncate the $H_\mathrm{HFI}$ at the magnetic-dipole
part
%===========================================================================
\begin{equation}
H_{\mathrm{HFI}} = \left( \frac{\mathbf{\mu}^{(1)}}{\mu_N}%
\cdot \mathcal{T}^{\left( 1\right) }\right) \, .
\label{Eq:HFI}
\end{equation}
%===========================================================================
Here $\mathbf{\mu }^{\left( 1\right) }$ is the operator of the nuclear
magnetic moment, $\mu_{N}$ is the nuclear magneton, and
$\mathcal{T}^{\left( 1\right) } $ is a relevant operator acting in the
electronic space.

While for the  $nsnp\,^{3}\!P_{0}$ the truncation of the HFI
Hamiltonian at the nuclear magnetic-dipole
contribution~(\ref{Eq:HFI}) is rigorously justified due to the
selection rules, for the $nsnp\,^{3}\!P_{2}$ states such a
truncation requires a special consideration. Indeed, the next
leading term in the multipole expansion of $H_{\mathrm{HFI}}$ is
due to the nuclear electric quadrupole moment. The associated
electronic tensor of rank 2 can also admix levels of $J=1$
symmetry and contribute to the hyperfine quenching. This
quadrupole contribution does not vanish for isotopes with $I \ge
1$, i.e.,  it may modify the $nsnp\,^{3}\!P_{2}$ for all the
isotopes listed in the Table~\ref{Tab:NucParams}, except for
$^{171}$Yb. Our truncation of the HFI Hamiltonian will lead to a
small ``systematic'' error for the $nsnp\,^{3}\!P_{2}$ hyperfine
quenching rates and we will return to this point in the
conclusion.

With the magnetic-dipole contribution, Eq.~(\ref{Eq:HFI}), the
required mixing matrix element  in Eq.~(\ref{Eq:WFCorr}) is
\begin{eqnarray}
\lefteqn{\langle \gamma^{\prime }(IJ^{\prime })F^{\prime }M_{F}^{\prime }|H_{\mathrm{HFI}%
}|\gamma (IJ) FM_{F}\rangle =
\delta _{FF^{\prime }}\delta _{M_{F}M_{F}^{\prime }} }  \nonumber \\
&& \times \left( -1\right)
^{F+I+J^{\prime }}\frac{\mu }{\mu _{N}}\sqrt{\frac{\left( 2I+1\right) \left(
I+1\right) }{I}}\left\{
\begin{tabular}{lll}
$I$ & $I$ & $1$ \\
$J$ & $J^{\prime }$ & $F$%
\end{tabular}
\right\} \langle \gamma^{\prime }J^{\prime }||\mathcal{T}^{\left( 1\right)
}||\gamma J\rangle \, ,
\end{eqnarray}
where $\mu$ are the nuclear magnetic moments, compiled
in Table~\ref{Tab:NucParams}.

%===========================================================================
Given the correction to the wavefunction, Eq.(\ref{Eq:WFCorr}),
we derive the hyperfine quenching rate using the standard formalism.
The rate of spontaneous emission ($a\rightarrow b$) for an electric-dipole
radiation is
%===========================================================================
\begin{equation}
A_{a\rightarrow b}=\frac{4\alpha^3 }{3}
\omega_{ab}^{3}\left| \langle a|\mathbf{D}|b\rangle \right| ^{2},
\end{equation}
%===========================================================================
where $\alpha \approx 1/137$ is the fine structure constant,
$\omega_{ab}=E_a-E_b$ is the transition frequency, and
$\mathbf{D}$ is the electric-dipole operator. Summing over all
possible $F_{b}$ and magnetic quantum numbers $M_{b}$ of the final
state, while disregarding small $F$-dependent energy correction,
one obtains
%===========================================================================
\begin{equation}
A_{a\rightarrow b}=\frac{4\alpha^3 }{3} \, \omega_{ab}^{3} \, %
\frac{1}{2F_{a}+1} \sum_{F_{b}} \left| \langle a||D||b\rangle \right| ^{2}.
\end{equation}
For the case at hand, the initial state is the HFI-perturbed
$nsnp\,^3\!P_J, (J \ne 1)$ state, and the final state is the ground
$ns^2\,^1\!S_0$ state.
Taking into account Eq.(\ref{Eq:WFCorr}), we arrive at the hyperfine quenching rate
%===========================================================================
\begin{equation}
A_{\mathrm{HFI}}\left( nsnp\,^3\!P_J;F\rightarrow ns^2\,^1\!S_0\right) =
\frac{4\alpha ^{3}}{9} \omega_J^3
\left( \frac{\mu }{\mu _{N}}\right)^{2}\frac{\left( 2I+1\right) \left(
I+1\right) }{I}\left\{
\begin{tabular}{lll}
$I$ & $I$ & $1$ \\
$J$ & $1$ & $F$%
\end{tabular}
\right\} ^{2} \, \left| S_J \right| ^{2} \, ,
\label{Ahfs}
\end{equation}
%===========================================================================
with
$\omega_J = E\left(nsnp\,^3\!P_J \right) -E\left( ns^2\,^1\!S_0 \right)$ and
the sums $S_J$   defined as
%===========================================================================
\begin{equation}
S_J=\sum_{\gamma^{\prime}, J^{\prime }}
\frac{
\langle nsnp\,^3\!P_J ||
 \mathcal{T}^{\left( 1\right)}|| \gamma^{\prime }J^{\prime } \rangle
\langle \gamma^{\prime } J^{\prime} ||D||  ns^2\,^1\!S_0 \rangle
 }
{ E \left( \gamma^{\prime }J^{\prime }\right) -E\left( nsnp\,^3\!P_J\right) }.
\label{S}
\end{equation}
%===========================================================================
Notice that due to the electric-dipole selection rules,
$J^\prime=1$. Also while the rate $A_{\mathrm{HFI}}$ depends on
the nuclear parameters and the value of $F$, the sums $S_J$ do
not.  In a particular case of the $nsnp\,^{3}\!P_{0}$ states, the
rate formula (\ref{Ahfs})  may be simplified to
\begin{equation}
A_{\mathrm{HFI}}\left( nsnp\,^3\!P_0;F=I\rightarrow ns^2\,^1\!S_0\right) =
\frac{4\alpha ^{3}}{27} \omega_0^3
\left( \frac{\mu }{\mu _{N}}\right)^{2}\,
\frac{ I+1 }{I} \, \left|S_0\right|^{2}.
\label{Eq:3P_0}
\end{equation}
In the following section we describe the {\em ab initio} relativistic
many-body calculations of the  derived hyperfine quenching rates.
%================================
\section{Solving atomic many-body problem}
%================================
\label{Sec:Method} The {\em ab initio} relativistic
atomic-structure calculations employed here are similar to
computations of electric-dipole amplitudes for the alkaline-earth
atoms \cite{PorKozRak01} and hyperfine structure constants and
electric-dipole amplitudes for ytterbium
\cite{PorRakKoz99J,PorRakKoz99P}. Here we only briefly recap the
main features of this method. We consider Mg, Ca, Sr and Yb as
atoms with two valence electrons outside the closed-shell cores.
Strong repulsion between the two valence electrons is treated
non-perturbatevely using the configuration-interaction (CI)
method. The core-valence and core-core correlations are taken into
account with the help of the many-body perturbation theory (MBPT)
method. In the following we refer to this combined approach as the
CI+MBPT method~\cite{DzuFlaKoz96b}.

In the CI+MBPT approach, the energies and the wave functions are
determined from the eigenvalue equation in the model space of the
valence electrons
%===========================================================================
\begin{equation}
H_{\mathrm{eff}}(E_{p})\,|\Phi _{p}\rangle =E_{p}\,|\Phi _{p}\rangle \,,
\label{Eqn_Sh}
\end{equation}
%===========================================================================
where the effective Hamiltonian is defined as
%===========================================================================
\begin{equation}
H_{\mathrm{eff}}(E)=H_{\mathrm{FC}}+\Sigma (E).
\label{Eqn_Heff}
\end{equation}
%===========================================================================
Here $H_{\rm{FC}}$ is the relativistic two-electron  Hamiltonian
in the frozen core approximation and $\Sigma(E)$ is the
energy-dependent core-polarization correction. The all-order
operator $\Sigma(E)$ completely accounts for the second order
correlation correction to the energies. The omitted diagrams in
higher orders may be accounted for indirectly by adjusting the
effective Hamiltonian~\cite{KozPor99O,PorKozRak01}. Namely, one
introduces an energy shift $\delta$ and replaces $\Sigma (E)$ with
$\Sigma (E-\delta )$. The parameter $\delta$ is determined
semi-empirically from a fit of the resulting theoretical energy
levels to experimental spectrum.

Using the effective Hamiltonian we find the wave functions of the
ground and the $^3\!P_J$ states. Further we apply the technique of
effective all-order (``dressed'') operators to calculations of the
matrix elements. Technically, we employ  the random-phase
approximation (RPA). The RPA sequence of diagrams describes a
shielding of externally applied field by the core electrons. This
is the level of approximation employed here for electric-dipole
matrix elements. The hyperfine, $\mathcal{T}^{(1)}$, matrix
elements required more sophisticated approach: for this operator
we additionally incorporated smaller corrections (e.g.,
normalization and structural radiation; the details can be found
in Ref.~\cite{PorRakKoz99J}). For the heaviest and more
computationally demanding Yb, the corrections to the effective
hyperfine operator tend to cancel~\cite{PorRakKoz99J}, and we have
simplified the calculations for Yb by using the bare
$\mathcal{T}^{(1)}$ operator.

To demonstrate the quality of the constructed wave functions and
the accuracy of the effective-operator approach, in
Table~\ref{Tab:hfs} we present the calculated magnetic-dipole
hyperfine structure constants $A$ for the $^3\!P_{1,2}$ states.
These constants are expressed in terms of expectation values of
$H_\mathrm{HFI}$. As seen from the Table~\ref{Tab:hfs} the
differences between the calculated and the experimental values,
even for heavy Yb, do not exceed 1\%.
%===========================================================================
\begin{table}
\caption{Magnetic-dipole hyperfine structure constants $A$ for the
$nsnp\,^{3}\!P_{1}$ and $nsnp\,^{3}\!P_2$ states.
The computed values are  compared
with the experimental data.
\label{Tab:hfs}}
\begin{ruledtabular}
\begin{tabular}{llll}
           &            & $A(^3P_1^o)$ (MHz)            & $A(^3P_2^o)$ (MHz)           \\
\hline
$^{25}$Mg  & This work  & -146.1                        & -129.7                       \\
           & Experiment & -144.977(5)  \footnotemark[1] & -128.445(5) \footnotemark[1] \\
$^{43}$Ca  & This work  & -199.2                        & -173.1                       \\
           & Experiment & -198.890(1)  \footnotemark[2] & -171.962(2) \footnotemark[3] \\
$^{87}$Sr  & This work  & -258.7                        & -211.4                       \\
           & Experiment & -260.083(5)  \footnotemark[4] & -212.765(1) \footnotemark[4] \\
$^{171}$Yb & This work  &  3964                         &  2704                        \\
           & Experiment &  3957.97(47) \footnotemark[5] &  2677.6     \footnotemark[6] \\
$^{173}$Yb & This work  & -1092                         & -745                          \\
           & Experiment & -1094.20(60) \footnotemark[5] & -737.7      \footnotemark[6]  \\
\end{tabular}
\end{ruledtabular}
\small \footnotemark[1]{\citet{Lur62},}
\small \footnotemark[2]{\citet{ArnBerBop81},}
\small \footnotemark[3]{\citet{GruGusLin79},}
\small \footnotemark[4]{\citet{HeiBri77},}
\small \footnotemark[5]{\citet{ClaCagLew79},}
\small \footnotemark[6]{\citet{BudSni69}.}
\end{table}
%===========================================================================
Further the sums $S_J$, Eq.~(\ref{S}), are computed in the framework of
Sternheimer-Dalgarno-Lewis  method \cite{Ste50,DalLew55}.
At the heart of this method is the recasting of the
sums  $S_J$ in the form
%===========================================================================
\begin{equation}
S_J =\langle nsnp\,^3\!P_J ||\mathcal{T}^{(1)}_\mathrm{eff}|| \delta\Psi   \rangle,
\label{T1}
\end{equation}
%===========================================================================
where $|\delta \Psi  \rangle$ satisfies the inhomogeneous Schrodinger
equation
%===========================================================================
\begin{equation}
\left( H_{\mathrm{eff}}-E(nsnp\,^3\!P_J) \right) |\delta \Psi \rangle
=  D_\mathrm{eff} |ns^2\, ^1\!S_0 \rangle \, .
\label{Eq:InhomSE}
\end{equation}
%===========================================================================
It is worth noting that because the effective operators act
in the valence model space, the $|\delta \Psi \rangle$ solution
encompasses only the excitations of the valence electrons to higher
valence states. The unaccounted for core excitations involve large energy
denominators and we disregard their contributions.

%===============================
\section{Results and conclusions}
%===============================
\label{Sec:Conclusions} To reiterate the discussion of the
previous section, we carry out the calculations in several logical
steps. First, we solve the CI+MBPT eigenvalue problem
(\ref{Eqn_Sh}) and determine the ground and the $nsnp\,^3\!P_J$
state wavefunctions and energies. At the next step, we compute the
dressed E1-operator $D_\mathrm{eff}$ and solve the inhomogeneous
equation (\ref{Eq:InhomSE}). Finally, we calculate the required
sums $S_J$, Eqs.(\ref{S}),(\ref{T1}), and determine the hyperfine
quenching  rates, Eq.~(\ref{Ahfs}).

The computed values of the isotope-independent sums $S_0$ and
$S_2$ for Mg, Ca, Sr, and Yb are presented in
Table~\ref{Tab:S}. The sums grow larger for heavier atoms
due to increasing matrix elements of the hyperfine interaction
(see Table~\ref{Tab:hfs}). A direct investigation of the
sums shows that the contributions of both $nsnp\,^3\!P_1$ and
$nsnp\,^1\!P_1$ intermediate states are comparable.
The triplet state is separated by just a fine-structure interval
from the metastable states, but its E1 matrix element with the singlet
ground state vanishes non-relativistically. For the singlet state, the
situation is reversed: compared to the triplet contribution,
the involved energy denominator is much larger, but the
electric-dipole matrix element is allowed.
%===========================================================================
\begin{table}[h]
\caption{Sums $S_J$ for the metastable $^{3}P_0$ and $^{3}P_2$
states. The values are given in atomic units.}
\label{Tab:S}
\begin{ruledtabular}
\begin{tabular}{ccc}
   &
\multicolumn{1}{c}{$S_0, nsnp\,^{3}\!P_0$}&
\multicolumn{1}{c}{$S_2, nsnp\,^{3}\!P_2$}     \\
\hline
Mg &   1.36 $\times 10^{-5}$ &   2.30$ \times 10^{-5}$  \\
Ca &   3.56 $\times 10^{-5}$ &   5.13$ \times 10^{-5}$  \\
Sr &   8.86 $\times 10^{-5}$ &   1.27$ \times 10^{-4}$  \\
Yb &   2.27 $\times 10^{-4}$ &   3.83$ \times 10^{-4}$  \\
\end{tabular}
\end{ruledtabular}
\end{table}
%===========================================================================

With the determined values of $S_J$ and Eq.(\ref{Ahfs}), we obtain
the hyperfine quenching rates for the metastable $^{3}P_{0}$ and
$^{3}P_{2}$ states. The resulting rates are listed in
Table~\ref{Tab:rates}. The tabulated decay rates for the
$^{3}P_{2}$ states require some explanation. First of all,  as
follows from Eq.(\ref{Ahfs}) the quenching rates depend on the
total angular momentum $F$ of the hyperfine substate. Although, in
general, the total angular momentum $F$ ranges from $|J-I|$ to
$J+I$, the 6j-symbol in Eq.(\ref{Ahfs}) imposes a stronger
restriction,  $|I-1| \le F \le I+1$. This requirement can be
tracked to the selection rule for the electric-dipole transition
amplitude between the ground state ($J_g=0,F_g=I$) and the
intermediate state which has the same  $F$ as the original
hyperfine state (see Eq.~(\ref{Eq:WFCorr})). Keeping this
restriction in mind, in Table~\ref{Tab:rates} we have listed the
quenching rates only for such E1-allowed values of $F$.

\begin{table}[h]
\caption{The hyperfine E1-quenching rates for the metastable
$^{3}P_0$ and $^{3}P_2$ states in sec$^{-1}$. The rates depend on
the total angular momentum $F$. The rates are compared with values
by other authors, where available.}
\label{Tab:rates}
\begin{ruledtabular}
\begin{tabular}{ccccc}
Atom & Transition rate & $F$ & This work & Other  \\
\hline
$^{25}$Mg & $A_{\rm{HFS}}(^{3}P_{0} \rightarrow \ ^{1}S_{0})$ & 5/2 &
$4.44\times 10^{-4}$ & $4.2\times 10^{-4}$ \footnotemark[1] \\
& $A_{\mathrm{HFS}}(^{3}P_{2} \rightarrow \ ^{1}S_{0})$ & 3/2 & $%
2.31\times 10^{-4}$ & $1.4\times 10^{-4}$ \footnotemark[1] \\
&  & 5/2 & $4.69\times 10^{-4}$ & $2.9\times 10^{-4}$ \footnotemark[1] \\
&  & 7/2 & $4.95\times 10^{-4}$ & $3.1\times 10^{-4}$ \footnotemark[1] \\
$^{43}$Ca & $A_{\mathrm{HFS}}(^{3}P_{0} \rightarrow \ ^{1}S_{0})$ & 7/2 &
$2.22\times 10^{-3}$ &  \\
& $A_{\mathrm{HFS}}(^{3}P_{2} \rightarrow \ ^{1}S_{0})$ & 5/2 & $%
1.02\times 10^{-3}$ &  \\
&  & 7/2 & $1.81\times 10^{-3}$ &  \\
&  & 9/2 & $1.74\times 10^{-3}$ &  \\
$^{87}$Sr & $A_{\mathrm{HFS}}(^{3}P_{0} \rightarrow \ ^{1}S_{0})$ & 9/2 &
$7.58\times 10^{-3}$ &  $6.3\times 10^{-3}$ \footnotemark[2] \\
& $A_{\mathrm{HFS}}(^{3}P_{2} \rightarrow \ ^{1}S_{0})$ & 7/2 & $%
4.13\times 10^{-3}$ &  \\
&  & 9/2 & $6.86\times 10^{-3}$ &  \\
&  & 11/2 & $6.27\times 10^{-3}$ &  \\
$^{171}$Yb & $A_{\mathrm{HFS}}(^{3}P_{0} \rightarrow \ ^{1}S_{0})$ & 1/2
& $4.35\times 10^{-2}$ &  $5.0 \times 10^{-2}$ \footnotemark[3] \\
& $A_{\mathrm{HFS}}(^{3}P_{2} \rightarrow \ ^{1}S_{0})$ & 3/2 & $%
9.18 \times 10^{-2}$ &   \\
$^{173}$Yb & $A_{\mathrm{HFS}}(^{3}P_{0} \rightarrow \ ^{1}S_{0})$ & 5/2
& $3.85\times 10^{-2}$ &  $4.3 \times 10^{-2}$ \footnotemark[3] \\
\end{tabular}
\end{ruledtabular}
\small \footnotemark[1]{\citet{Gar62},}
\small \footnotemark[2]{\citet{KatTakPal03},}
\small \footnotemark[2]{\citet{PorDerFor03}.}
\end{table}
%===========================================================================

We also remind the reader that in our analysis we have disregarded
the contributions of the quadrupole and higher-order nuclear
magnetic moments. While for the $J=0$ states this truncation is
rigorously justified, for the $J=2$ states the quadrupole
contribution is generally present and becomes increasingly
important for heavier atoms. Its relative role may be roughly
estimated by forming a ratio of the relevant electric-quadrupole,
$B$, and the magnetic-dipole, $A$, hyperfine-structure constants.
The ratio $B/A$ for the $nsnp\,^{3}\!P_{2}$ states is less or
in the order of 0.1 for isotopes of Mg, Ca, and Sr, but is larger
than unity for $^{173}$Yb. We expect that the quadrupole
correction will be significant for $^{173}$Yb. However, for the
$^{171}$Yb isotope, $I=1/2$ and the nuclear moments beyond the
magnetic-dipole moment vanish, justifying the validity of the
truncation.

Based on better than 1\% accuracy of the {\em ab initio} hyperfine
constants (Table~\ref{Tab:hfs}) and energy
levels~\cite{PorKozRak01,PorRakKoz99J} we expect that the computed
hyperfine quenching rates for the $^3\!P_0$ states are accurate
within at least a few per cent. For the $^3\!P_2$ states of
alkaline-earth atoms the main source of uncertainty is due to the
neglected nuclear quadrupole moment contributions; the overall
accuracy should be worse than that.

In Table~\ref{Tab:rates} we also compare the computed rates with
the results from the literature. For Mg the hyperfine quenching
rates for the $^{3}\!P_2$ state were estimated more than four
decades ago by \citet{Gar62}. Our results are in a reasonable
agreement with his values. Certainly, our calculations based on
the modern {\em ab initio} relativistic many-body techniques are
more complete. For instance, in the calculation of the sums $S_J$,
\citet{Gar62} kept only the two lowest-energy intermediate states
$^3\!P_1$ and $^1\!P_1$. This author has also employed the
following E1-matrix elements $ |\langle ^1S_0||D||^3P_1 \rangle| =
0.0058$ a.u. and $ |\langle ^1S_0||D||^1P_1 \rangle| = 3.46$ a.u.,
which are smaller than more accurate values~\cite{PorKozRak01} of
0.0064(7) a.u. and 4.03(2) a.u., employed here. Our results for
$^{87}$Sr  are in fair agreement with the estimate of
Ref.~\cite{KatTakPal03}. Previously, we have estimated the
quenching rates for Yb isotopes~\cite{PorDerFor03} by summing only
over the two lowest-energy excited states; the present result
should be considered as more accurate.

In Table~\ref{hfsM2} the calculated hyperfine quenching rates
(maximum over hyperfine manifold) for the $^3\!P_2$ states are
compared with the conventional electromagnetic transition rates.
For Mg, Ca, and Sr these rates were calculated in
Ref.~\cite{Der01} and are due to M1, M2, E2, and E3 multipole
transitions. If the hyperfine quenching is allowed for a
particular value of $F$, both rates  contribute at a comparable
level for Mg. For Ca and heavier atoms the hyperfine quenching
becomes the dominant decay branch and determines the lifetime of
the fermionic isotopes.

It is worth mentioning that \citet{YasKat03} have experimentally
demonstrated that the blackbody radiation at 300 K quenches the
$^3\!P_J$ metastable states of Sr, significantly shortening their
lifetimes. We verified that for other atoms, Mg, Ca, and Yb at $T
< 300$ K the blackbody radiation does not affect the lifetimes of
the $^3\!P_J$ states. We would like to emphasize, that the results
tabulated in this paper are for the radiative decay rates due to
the vacuum fluctuations of the electromagnetic field, i.e., for
the ambient temperature of $T=0$. The additional quenching by the
temperature-dependent blackbody radiation~\cite{YasKat03} should
be also included in the total rate, especially for Sr isotopes.

\begin{table}[h]
\caption{Comparison of the  hyperfine quenching rates (maximum
over the hyperfine manifold in Table~\protect\ref{Tab:rates}) with
the non-dipole rates for the $^3P_2$ states. The rates are given
in sec$^{-1}$.}
\label{hfsM2}
\begin{ruledtabular}
\begin{tabular}{ccc}
    Atom    & hyperfine rate, max  &   non-E1 rate \footnotemark[1] \\
\hline
 $^{25}$Mg  &  4.95 $\times 10^{-4}$  &   4.42 $\times 10^{-4}$  \\
 $^{43}$Ca  &  18.1 $\times 10^{-4}$ &  1.41 $\times 10^{-4}$  \\
 $^{87}$Sr  &  68.6 $\times 10^{-4}$  &   9.55 $\times 10^{-4}$  \\
$^{171}$Yb  &  9.18 $\times 10^{-2}$  &  6.7  $\times 10^{-2}$  \\
\end{tabular}
\end{ruledtabular}
\small \footnotemark[1]{For Mg, Ca, and Sr the rates are from Ref.~\cite{Der01}
and for Yb from Ref.~\cite{PorRakKoz99P}.}
\end{table}

To summarize, here we employed the relativistic many-body
methods to evaluate the hyperfine quenching rates for
the metastable $^{3}\!P_{0}$ and $^{3}\!P_{2}$ states
of Mg, Ca, Sr, and Yb. The tabulated rates may be useful
for designing  novel ultra-precise optical clocks
and trapping experiments with fermionic
isotopes of metastable alkaline-earth atoms and Yb.
The resulting natural widths of the $^3\!P_0-^1\!S_0$
clock transition are 0.44 mHz for $^{25}$Mg,
2.2 mHz for $^{43}$Ca, 7.6 mHz for $^{87}$Sr, 43.5 mHz for
$^{171}$Yb, and 38.5 mHz for $^{173}$Yb. Compared to
the bosonic isotopes, the lifetime of the
$^{3}\!P_{2}$ states in fermionic isotopes is noticeably  shortened
by the hyperfine quenching but still remains long enough for trapping
experiments.

%========================
\acknowledgments
We would like to thank Norval Fortson for bringing this problem
to our attention.
This work was supported in part by the National Science
Foundation grant and by the NIST  Precision Measurement grant.
The work of S.G.P. was
additionally supported by the Russian Foundation for Basic
Research under grant No. 02-02-16837-a.

%===========================================================================
%\bibliographystyle{plain}
%\bibliography{hfs,Por_ref,clocks,Der_ref,scat3P2_add,all}

\begin{thebibliography}{29}
\expandafter\ifx\csname natexlab\endcsname\relax\def\natexlab#1{#1}\fi
\expandafter\ifx\csname bibnamefont\endcsname\relax
  \def\bibnamefont#1{#1}\fi
\expandafter\ifx\csname bibfnamefont\endcsname\relax
  \def\bibfnamefont#1{#1}\fi
\expandafter\ifx\csname citenamefont\endcsname\relax
  \def\citenamefont#1{#1}\fi
\expandafter\ifx\csname url\endcsname\relax
  \def\url#1{\texttt{#1}}\fi
\expandafter\ifx\csname urlprefix\endcsname\relax\def\urlprefix{URL }\fi
\providecommand{\bibinfo}[2]{#2}
\providecommand{\eprint}[2][]{\url{#2}}

\bibitem[{Cop()}]{CopenhagenWorkshop2003}
\bibinfo{note}{See, e.g., abstracts of the Second Workshop on Cold
  Alkaline-Earth Atoms, held September 11-13, 2003 in Copenhagen, Denmark.
  Abstracts are available from
  http://www.fys.ku.dk/coldatoms/workshop/workshopgroup2.htm}.

\bibitem[{\citenamefont{Takasu et~al.}(2003)\citenamefont{Takasu, Maki, Komori,
  Takano, Honda, Kumakura, Yabuzaki, and Takahashi}}]{TakMakKom03}
\bibinfo{author}{\bibfnamefont{Y.}~\bibnamefont{Takasu}},
  \bibinfo{author}{\bibfnamefont{K.}~\bibnamefont{Maki}},
  \bibinfo{author}{\bibfnamefont{K.}~\bibnamefont{Komori}},
  \bibinfo{author}{\bibfnamefont{T.}~\bibnamefont{Takano}},
  \bibinfo{author}{\bibfnamefont{K.}~\bibnamefont{Honda}},
  \bibinfo{author}{\bibfnamefont{M.}~\bibnamefont{Kumakura}},
  \bibinfo{author}{\bibfnamefont{T.}~\bibnamefont{Yabuzaki}}, \bibnamefont{and}
  \bibinfo{author}{\bibfnamefont{Y.}~\bibnamefont{Takahashi}},
  \bibinfo{journal}{Phys. Rev. Lett.} \textbf{\bibinfo{volume}{91}},
  \bibinfo{pages}{040404} (\bibinfo{year}{2003}).

\bibitem[{\citenamefont{Katori et~al.}(2001)\citenamefont{Katori, Ido, Isoya,
  and Kuwata-Gonokami}}]{KatIdoIso00}
\bibinfo{author}{\bibfnamefont{H.}~\bibnamefont{Katori}},
  \bibinfo{author}{\bibfnamefont{T.}~\bibnamefont{Ido}},
  \bibinfo{author}{\bibfnamefont{Y.}~\bibnamefont{Isoya}}, \bibnamefont{and}
  \bibinfo{author}{\bibfnamefont{M.}~\bibnamefont{Kuwata-Gonokami}}, in
  \emph{\bibinfo{booktitle}{Atomic Physics 17}}, edited by
  \bibinfo{editor}{\bibfnamefont{E.}~\bibnamefont{Arimondo}},
  \bibinfo{editor}{\bibfnamefont{P.}~\bibnamefont{DeNatale}}, \bibnamefont{and}
  \bibinfo{editor}{\bibfnamefont{M.}~\bibnamefont{Inguscio}}
  (\bibinfo{publisher}{AIP, New York}, \bibinfo{year}{2001}).

\bibitem[{\citenamefont{Nagel et~al.}(2003)\citenamefont{Nagel, Simien, Laha,
  Gupta, Ashoka, and Killian}}]{NagSimLah03}
\bibinfo{author}{\bibfnamefont{S.~B.} \bibnamefont{Nagel}},
  \bibinfo{author}{\bibfnamefont{C.~E.} \bibnamefont{Simien}},
  \bibinfo{author}{\bibfnamefont{S.}~\bibnamefont{Laha}},
  \bibinfo{author}{\bibfnamefont{P.}~\bibnamefont{Gupta}},
  \bibinfo{author}{\bibfnamefont{V.~S.} \bibnamefont{Ashoka}},
  \bibnamefont{and} \bibinfo{author}{\bibfnamefont{T.~C.}
  \bibnamefont{Killian}}, \bibinfo{journal}{Phys. Rev. A}
  \textbf{\bibinfo{volume}{67}}, \bibinfo{pages}{011401(R)}
  (\bibinfo{year}{2003}).

\bibitem[{\citenamefont{Xu et~al.}(2003)\citenamefont{Xu, Loftus, Hall,
  Gallagher, and Ye}}]{XuLofHal03}
\bibinfo{author}{\bibfnamefont{X.}~\bibnamefont{Xu}},
  \bibinfo{author}{\bibfnamefont{T.~H.} \bibnamefont{Loftus}},
  \bibinfo{author}{\bibfnamefont{J.~L.} \bibnamefont{Hall}},
  \bibinfo{author}{\bibfnamefont{A.}~\bibnamefont{Gallagher}},
  \bibnamefont{and} \bibinfo{author}{\bibfnamefont{J.}~\bibnamefont{Ye}},
  \bibinfo{journal}{J.\ Opt.\ Soc.\ Am.\ B} \textbf{\bibinfo{volume}{20}},
  \bibinfo{pages}{968} (\bibinfo{year}{2003}).

\bibitem[{\citenamefont{Gr\"{u}nert and Hemmerich}(2002)}]{GruHem02}
\bibinfo{author}{\bibfnamefont{J.}~\bibnamefont{Gr\"{u}nert}} \bibnamefont{and}
  \bibinfo{author}{\bibfnamefont{A.}~\bibnamefont{Hemmerich}},
  \bibinfo{journal}{Phys. Rev. A} \textbf{\bibinfo{volume}{65}},
  \bibinfo{pages}{041401} (\bibinfo{year}{2002}).

\bibitem[{\citenamefont{Katori et~al.}(2003)\citenamefont{Katori, Takamoto,
  Pal'chikov, and Ovsiannikov}}]{KatTakPal03}
\bibinfo{author}{\bibfnamefont{H.}~\bibnamefont{Katori}},
  \bibinfo{author}{\bibfnamefont{M.}~\bibnamefont{Takamoto}},
  \bibinfo{author}{\bibfnamefont{V.~G.} \bibnamefont{Pal'chikov}},
  \bibnamefont{and} \bibinfo{author}{\bibfnamefont{V.~D.}
  \bibnamefont{Ovsiannikov}}, \bibinfo{journal}{Phys.\ Rev.\ Lett.}
  \textbf{\bibinfo{volume}{91}}, \bibinfo{pages}{173005}
  (\bibinfo{year}{2003}).

\bibitem[{\citenamefont{Courtillot et~al.}(2003)\citenamefont{Courtillot,
  Quessada, Kovacich, Brusch, Kolker, Zondy, Rovera, and
  Lemonde}}]{CouQueKov03}
\bibinfo{author}{\bibfnamefont{I.}~\bibnamefont{Courtillot}},
  \bibinfo{author}{\bibfnamefont{A.}~\bibnamefont{Quessada}},
  \bibinfo{author}{\bibfnamefont{R.~P.} \bibnamefont{Kovacich}},
  \bibinfo{author}{\bibfnamefont{A.}~\bibnamefont{Brusch}},
  \bibinfo{author}{\bibfnamefont{D.}~\bibnamefont{Kolker}},
  \bibinfo{author}{\bibfnamefont{J.-J.} \bibnamefont{Zondy}},
  \bibinfo{author}{\bibfnamefont{G.~D.} \bibnamefont{Rovera}},
  \bibnamefont{and} \bibinfo{author}{\bibfnamefont{P.}~\bibnamefont{Lemonde}},
  \bibinfo{journal}{Phys.\ Rev. A} \textbf{\bibinfo{volume}{68}},
  \bibinfo{pages}{030501(R)} (\bibinfo{year}{2003}).

\bibitem[{\citenamefont{Park and Yoon}(2003)}]{ParYoo03}
\bibinfo{author}{\bibfnamefont{C.~Y.} \bibnamefont{Park}} \bibnamefont{and}
  \bibinfo{author}{\bibfnamefont{T.~H.} \bibnamefont{Yoon}},
  \bibinfo{journal}{Phys.\ Rev.\ A} \textbf{\bibinfo{volume}{68}},
  \bibinfo{pages}{055401} (\bibinfo{year}{2003}).

\bibitem[{\citenamefont{Porsev et~al.}(2003)\citenamefont{Porsev, Derevianko,
  and Fortson}}]{PorDerFor03}
\bibinfo{author}{\bibfnamefont{S.~G.} \bibnamefont{Porsev}},
  \bibinfo{author}{\bibfnamefont{A.}~\bibnamefont{Derevianko}},
  \bibnamefont{and} \bibinfo{author}{\bibfnamefont{E.~N.}
  \bibnamefont{Fortson}} (\bibinfo{year}{2003}), \bibinfo{note}{submitted to
  Phys. Rev. A}.

\bibitem[{\citenamefont{Porsev et~al.}(1999{\natexlab{a}})\citenamefont{Porsev,
  Rakhlina, and Kozlov}}]{PorRakKoz99P}
\bibinfo{author}{\bibfnamefont{S.~G.} \bibnamefont{Porsev}},
  \bibinfo{author}{\bibfnamefont{{\rm Yu}.~G.} \bibnamefont{Rakhlina}},
  \bibnamefont{and} \bibinfo{author}{\bibfnamefont{M.~G.}
  \bibnamefont{Kozlov}}, \bibinfo{journal}{Phys. Rev. A}
  \textbf{\bibinfo{volume}{60}}, \bibinfo{pages}{2781}
  (\bibinfo{year}{1999}{\natexlab{a}}).

\bibitem[{\citenamefont{Derevianko}(2001)}]{Der01}
\bibinfo{author}{\bibfnamefont{A.}~\bibnamefont{Derevianko}},
  \bibinfo{journal}{Phys.\ Rev.\ Lett.} \textbf{\bibinfo{volume}{87}},
  \bibinfo{pages}{023002} (\bibinfo{year}{2001}).

\bibitem[{\citenamefont{Garstang}(1962)}]{Gar62}
\bibinfo{author}{\bibfnamefont{R.~G.} \bibnamefont{Garstang}},
  \bibinfo{journal}{J. Opt. Soc. Am.} \textbf{\bibinfo{volume}{52}},
  \bibinfo{pages}{845} (\bibinfo{year}{1962}).

\bibitem[{\citenamefont{Johnson et~al.}(1997)\citenamefont{Johnson, Cheng, and
  Plante}}]{JohChePla97}
\bibinfo{author}{\bibfnamefont{W.~R.} \bibnamefont{Johnson}},
  \bibinfo{author}{\bibfnamefont{K.~T.} \bibnamefont{Cheng}}, \bibnamefont{and}
  \bibinfo{author}{\bibfnamefont{D.~R.} \bibnamefont{Plante}},
  \bibinfo{journal}{Phys.\ Rev.\ A} \textbf{\bibinfo{volume}{55}},
  \bibinfo{pages}{2728} (\bibinfo{year}{1997}).

\bibitem[{\citenamefont{Brage et~al.}(1998)\citenamefont{Brage, Judge,
  Aboussaid, Godefroid, Jonsson, Ynnerman, Fischer, and
  Leckrone}}]{BraJudAbo98}
\bibinfo{author}{\bibfnamefont{T.}~\bibnamefont{Brage}},
  \bibinfo{author}{\bibfnamefont{P.~G.} \bibnamefont{Judge}},
  \bibinfo{author}{\bibfnamefont{A.}~\bibnamefont{Aboussaid}},
  \bibinfo{author}{\bibfnamefont{M.~R.} \bibnamefont{Godefroid}},
  \bibinfo{author}{\bibfnamefont{P.}~\bibnamefont{Jonsson}},
  \bibinfo{author}{\bibfnamefont{A.}~\bibnamefont{Ynnerman}},
  \bibinfo{author}{\bibfnamefont{C.~F.} \bibnamefont{Fischer}},
  \bibnamefont{and} \bibinfo{author}{\bibfnamefont{D.~S.}
  \bibnamefont{Leckrone}}, \bibinfo{journal}{Astrophys.\ J.}
  \textbf{\bibinfo{volume}{500}}, \bibinfo{pages}{507} (\bibinfo{year}{1998}).

\bibitem[{\citenamefont{Brage et~al.}(2002)\citenamefont{Brage, Judge, and
  Proffitt}}]{BraJudPro02}
\bibinfo{author}{\bibfnamefont{T.}~\bibnamefont{Brage}},
  \bibinfo{author}{\bibfnamefont{P.~G.} \bibnamefont{Judge}}, \bibnamefont{and}
  \bibinfo{author}{\bibfnamefont{C.~R.} \bibnamefont{Proffitt}},
  \bibinfo{journal}{Phys.\ Rev.\ Lett.} \textbf{\bibinfo{volume}{89}},
  \bibinfo{pages}{281101/1} (\bibinfo{year}{2002}).

\bibitem[{\citenamefont{Porsev et~al.}(2001)\citenamefont{Porsev, Kozlov,
  Rakhlina, and Derevianko}}]{PorKozRak01}
\bibinfo{author}{\bibfnamefont{S.~G.} \bibnamefont{Porsev}},
  \bibinfo{author}{\bibfnamefont{M.~G.} \bibnamefont{Kozlov}},
  \bibinfo{author}{\bibfnamefont{{\rm Yu}.~G.} \bibnamefont{Rakhlina}},
  \bibnamefont{and}
  \bibinfo{author}{\bibfnamefont{A.}~\bibnamefont{Derevianko}},
  \bibinfo{journal}{Phys. Rev. A} \textbf{\bibinfo{volume}{64}},
  \bibinfo{pages}{012508} (\bibinfo{year}{2001}).

\bibitem[{\citenamefont{Porsev et~al.}(1999{\natexlab{b}})\citenamefont{Porsev,
  Rakhlina, and Kozlov}}]{PorRakKoz99J}
\bibinfo{author}{\bibfnamefont{S.~G.} \bibnamefont{Porsev}},
  \bibinfo{author}{\bibfnamefont{{\rm Yu}.~G.} \bibnamefont{Rakhlina}},
  \bibnamefont{and} \bibinfo{author}{\bibfnamefont{M.~G.}
  \bibnamefont{Kozlov}}, \bibinfo{journal}{J. Phys. B}
  \textbf{\bibinfo{volume}{32}}, \bibinfo{pages}{1113}
  (\bibinfo{year}{1999}{\natexlab{b}}).

\bibitem[{\citenamefont{Dzuba et~al.}(1996)\citenamefont{Dzuba, Flambaum, and
  Kozlov}}]{DzuFlaKoz96b}
\bibinfo{author}{\bibfnamefont{V.~A.} \bibnamefont{Dzuba}},
  \bibinfo{author}{\bibfnamefont{V.~V.} \bibnamefont{Flambaum}},
  \bibnamefont{and} \bibinfo{author}{\bibfnamefont{M.~G.}
  \bibnamefont{Kozlov}}, \bibinfo{journal}{Phys.\ Rev.\ A}
  \textbf{\bibinfo{volume}{54}}, \bibinfo{pages}{3948} (\bibinfo{year}{1996}).

\bibitem[{\citenamefont{Kozlov and Porsev}(1999)}]{KozPor99O}
\bibinfo{author}{\bibfnamefont{M.~G.} \bibnamefont{Kozlov}} \bibnamefont{and}
  \bibinfo{author}{\bibfnamefont{S.~G.} \bibnamefont{Porsev}},
  \bibinfo{journal}{Opt. Spectrosk.} \textbf{\bibinfo{volume}{87}},
  \bibinfo{pages}{384} (\bibinfo{year}{1999}), \bibinfo{note}{[Opt. \
  Spectrosc. {\bf 87} 352, (1999)]}.

\bibitem[{\citenamefont{Lurio}(1962)}]{Lur62}
\bibinfo{author}{\bibfnamefont{A.}~\bibnamefont{Lurio}},
  \bibinfo{journal}{Phys. Rev.} \textbf{\bibinfo{volume}{126}},
  \bibinfo{pages}{1768} (\bibinfo{year}{1962}).

\bibitem[{\citenamefont{Arnold et~al.}(1981)\citenamefont{Arnold, Bergmann,
  Bopp, Dorsch, Kowalski, Stehlin, and Trager}}]{ArnBerBop81}
\bibinfo{author}{\bibfnamefont{M.}~\bibnamefont{Arnold}},
  \bibinfo{author}{\bibfnamefont{E.}~\bibnamefont{Bergmann}},
  \bibinfo{author}{\bibfnamefont{P.}~\bibnamefont{Bopp}},
  \bibinfo{author}{\bibfnamefont{C.}~\bibnamefont{Dorsch}},
  \bibinfo{author}{\bibfnamefont{J.}~\bibnamefont{Kowalski}},
  \bibinfo{author}{\bibfnamefont{T.}~\bibnamefont{Stehlin}}, \bibnamefont{and}
  \bibinfo{author}{\bibfnamefont{F.}~\bibnamefont{Trager}},
  \bibinfo{journal}{Hyperfine Int.} \textbf{\bibinfo{volume}{9}},
  \bibinfo{pages}{159} (\bibinfo{year}{1981}).

\bibitem[{\citenamefont{Grundevik et~al.}(1979)\citenamefont{Grundevik,
  Gustavsson, Lindgren, Olsson, Robertsson, Rosen, and Svanberg}}]{GruGusLin79}
\bibinfo{author}{\bibfnamefont{P.}~\bibnamefont{Grundevik}},
  \bibinfo{author}{\bibfnamefont{M.}~\bibnamefont{Gustavsson}},
  \bibinfo{author}{\bibfnamefont{I.}~\bibnamefont{Lindgren}},
  \bibinfo{author}{\bibfnamefont{G.}~\bibnamefont{Olsson}},
  \bibinfo{author}{\bibfnamefont{L.}~\bibnamefont{Robertsson}},
  \bibinfo{author}{\bibfnamefont{A.}~\bibnamefont{Rosen}}, \bibnamefont{and}
  \bibinfo{author}{\bibfnamefont{S.}~\bibnamefont{Svanberg}},
  \bibinfo{journal}{Phys. Rev. Lett.} \textbf{\bibinfo{volume}{42}},
  \bibinfo{pages}{1528} (\bibinfo{year}{1979}).

\bibitem[{\citenamefont{Heider and Brink}(1977)}]{HeiBri77}
\bibinfo{author}{\bibfnamefont{S.~M.} \bibnamefont{Heider}} \bibnamefont{and}
  \bibinfo{author}{\bibfnamefont{G.~O.} \bibnamefont{Brink}},
  \bibinfo{journal}{Phys. Rev. A} \textbf{\bibinfo{volume}{16}},
  \bibinfo{pages}{1371} (\bibinfo{year}{1977}).

\bibitem[{\citenamefont{Clark et~al.}(1979)\citenamefont{Clark, Cage, Lewis,
  and Greenlees}}]{ClaCagLew79}
\bibinfo{author}{\bibfnamefont{D.~L.} \bibnamefont{Clark}},
  \bibinfo{author}{\bibfnamefont{M.~E.} \bibnamefont{Cage}},
  \bibinfo{author}{\bibfnamefont{D.~A.} \bibnamefont{Lewis}}, \bibnamefont{and}
  \bibinfo{author}{\bibfnamefont{G.~W.} \bibnamefont{Greenlees}},
  \bibinfo{journal}{Phys. Rev. A} \textbf{\bibinfo{volume}{20}},
  \bibinfo{pages}{239} (\bibinfo{year}{1979}).

\bibitem[{\citenamefont{Budick and Snir}(1969)}]{BudSni69}
\bibinfo{author}{\bibfnamefont{B.}~\bibnamefont{Budick}} \bibnamefont{and}
  \bibinfo{author}{\bibfnamefont{J.}~\bibnamefont{Snir}},
  \bibinfo{journal}{Phys. Rev.} \textbf{\bibinfo{volume}{178}},
  \bibinfo{pages}{18} (\bibinfo{year}{1969}).

\bibitem[{\citenamefont{Sternheimer}(1950)}]{Ste50}
\bibinfo{author}{\bibfnamefont{R.~M.} \bibnamefont{Sternheimer}},
  \bibinfo{journal}{Phys. Rev.} \textbf{\bibinfo{volume}{80}},
  \bibinfo{pages}{102} (\bibinfo{year}{1950}).

\bibitem[{\citenamefont{Dalgarno and Lewis}(1955)}]{DalLew55}
\bibinfo{author}{\bibfnamefont{A.}~\bibnamefont{Dalgarno}} \bibnamefont{and}
  \bibinfo{author}{\bibfnamefont{J.~T.} \bibnamefont{Lewis}},
  \bibinfo{journal}{Proc. Roy. Soc.} \textbf{\bibinfo{volume}{223}},
  \bibinfo{pages}{70} (\bibinfo{year}{1955}).

\bibitem[{\citenamefont{Yasuda and Katori}(2003)}]{YasKat03}
\bibinfo{author}{\bibfnamefont{M.}~\bibnamefont{Yasuda}} \bibnamefont{and}
  \bibinfo{author}{\bibfnamefont{H.}~\bibnamefont{Katori}}
  (\bibinfo{year}{2003}), \bibinfo{note}{arXive:physics/0309044}.

\end{thebibliography}

\end{document}